\documentclass[aps,tighten,amssymb,twocolumn]{revtex4}
\usepackage{graphicx}

\newcommand{\C}{\mathbb{C}}
\newcommand{\N}{\mathbb{N}}
\newcommand{\R}{\mathbb{R}}
\newcommand{\Z}{\mathbb{Z}}
\newcommand{\BB}{\mathcal{B}}
\newcommand{\CC}{\mathcal{C}}
\newcommand{\II}{\mathcal{I}}
\newcommand{\KK}{\mathcal{K}}
\newcommand{\LL}{\mathcal{L}}
\newcommand{\NN}{\mathcal{N}}
\newcommand{\OO}{\mathcal{O}}
\newcommand{\PP}{\mathcal{P}}
\newcommand{\VV}{\mathcal{V}}
\newcommand{\XX}{\mathcal{X}}
\newcommand{\im}{\mathrm{Im\,}}

\begin{document}

\title{Approximations by graphs and emergence of global structures}

\author{Pavel Exner$^{1,2}$, Pavel Hej\v{c}\'{\i}k$^3$, and Petr \v{S}eba$^{2,3}$}
\address{
$^{1}$Nuclear Physics Institute, Czech Academy of Sciences,
25068 \v{R}e\v{z} near Prague, Czechia \\
$^{2}$Doppler Institute, Czech Technical University,
B\v{r}ehov\'{a} 7, 11519 Prague, Czechia \\
$^{3}$University of Hradec Kr\'alov\'{e}, V\'{\i}ta Nejed\'eho
573, 50002 Hradec Kr\'alov\'{e}, Czechia }

\date{\today}

\begin{abstract}
We study approximations of billiard systems by lattice graphs. It
is demonstrated that under natural assumptions the graph
wavefunctions approximate solutions of the Schr\"odinger equation
with energy rescaled by the billiard dimension. As an example, we
analyze a Sinai billiard with attached leads. The results
illustrate emergence of global structures in large quantum graphs
and offer interesting comparisons with patterns observed in
complex networks of a different nature.
\end{abstract}

\pacs{PACS: ???}

\maketitle

\narrowtext

\vspace*{-5mm}

\section{Introduction}

The notion of a quantum graph is known for more then half a
century \cite{RuS}, however, an intense investigation of these
structures started less than two decades ago \cite{ARZ, ES1} as a
response to progress in fabrication technologies which allowed to
prepare microscopic graph-like structures. Nowadays, there is an
extensive literature devoted to the subject; for recent reviews
see \cite{KS, Ku} and also \cite{AGHH}.

An attention to quantum graphs come from the fact that motion on
their edges is easy to describe, and at the same time the graph
structure leads to a nontrivial behavior. It was shown, in
particular, that even a graph with a small number of vertices is
capable of developing an internal dynamics rich enough to display
universality features that are typical for the wave-chaotic
behavior \cite{BBK, ks1, ks2}. It is not only a theory, the
results can be checked experimentally in a microwave graph model
\cite{sirko1}.

On the contrary, properties of nontrivial \emph{large-scale}
graphs have been regarded as less interesting due to the expected
localization of the corresponding wavefunctions. An indication
that this belief is wrong may be seen from the fact that complex
graph-like structures, such as systems of interconnected neurons,
display surprising patterns observed, for instance, in the visual
cortex of mammals \cite{blasdel}. It was shown in \cite{Geisel}
that these patterns can be understood as a manifestation of a
Gaussian random field, and are in this sense analogous to patterns
emerging in two dimensional quantum chaotic systems, for instance,
nodal domains in a chaotic quantum billiard \cite{blum}.

A thorough investigation of such structures on graphs is by no
means easy. To follow the mentioned example, a nodal domain is a
connected component of the maximal induced subgraph of a graph
$\Gamma$ on which a function does not change sign; it relates the
pattern formation on $\Gamma$ to nontrivial algebraic questions
about graph partition, etc. This is probably the reason why only
few mathematical results of this type are available at present -
cf.~\cite{biyi1, biyi2}.

In this paper we are going to show that extended graphs support
structures similar to those known from two-dimensional wave chaos
and Gaussian random-field models. We will show that the structures
do not depend on the local set up of the graph. They reflect the
influence of the graph boundary on the wave propagation and
interference on the graph. The final patterns are "global" in the
sense that they extend over hundreds of graph vertices.

Our approach is based on graph embedding into a Euclidean space
and a convergence argument; we will demonstrate that wavefunctions
on the graph can approximate solutions of the respective
``continuous'' billiard problem. The embedding assumption is
naturally a nontrivial restriction because not every graph can be
regarded as a subset of a Euclidean space from which it inherits
its metric. It applies, however, to wide enough class of systems
and allows us at the same time to circumvent difficulties of a
pure algebraic treatment.

The technical tool to derive the approximation result is a graph
duality adopted from \cite{Ex1}. To make the paper self-contained
we review this theory in the next section in a simplified form
suitable for the present purpose. Then we will show how solutions
to the Schr\"odinger equation can be approximated by those of a
Schr\"odinger equation on lattice graphs with the energy properly
rescaled.

To illustrate the result we will analyze an example of a lattice
graph which approximates a Sinai billiard. Since we want to go
beyond the nodal structure and to analyze also the phase behavior
of the wavefunctions we will study such a also from the transport
point of view, attaching to it a pair of semiinfinite leads; the
result will be compared to the "true" Sinai billiard with a pair
of leads attached. We will compare, in particular, the probability
currents and show they are similar to each other provided the
current on the graph is properly defined as a vector sum of
currents at the graphs links.


\setcounter{equation}{0}
\section{Theory: a graph duality}

By $\Gamma$ we denote in the following a connected graph
consisting of at most countable families of \emph{vertices}
$\VV=\{{\XX}_{j}:\,j\in I\}$ and \emph{edges} $\LL=\{\LL_{jn}:\,
(j,n)\in I_{\LL}\subset I\times I\}$. We suppose that each pair of
vertices is connected by not more than one link, otherwise we can
simply add vertices to any ``multiple'' edge. The set $\NN(\XX_j)=
\{\XX_n:\, n\in\nu(j)\subset I\setminus\{j\}\}$ consists of the
\emph{neighbors} of $\XX_j$, i.e. the vertices connected with
$\XX_j$ by a single edge is nonempty by assumption. The graph
\emph{boundary} $\BB$ consists of vertices having a single
neighbor; it may be empty. We denote by $I_\BB$ and $I_\II$ the
index subsets in $I$ corresponding to $\BB$ and the graph
\emph{interior} $\II:=\VV\setminus\BB$, respectively.

We suppose that $\Gamma$ is a \emph{metric graph}, i.e. that it
has a \emph{local} metric structure, every edge $\LL_{jn}$ being
isometric with a line segment $[0,\ell_{jn}]$. Of course, the
graph can be also equipped with a \emph{global} metric, for
instance, by identifying it with a subset of $\R^\nu$. In general
the metrics may not coincide, however, in the next section we will
identify them. Using the local metric, we introduce the Hilbert
space $L^2(\Gamma):= \bigoplus_{(j,n) \in I_\LL} L^2(0,\ell_{jn})$
whose elements are $\psi=\{\psi_{jn}:\, (j,n)\in I_\LL\}$ or
simply $\{\psi_{jn}\}$; in the same way we define Sobolev spaces
on $\Gamma$. Given a family of potentials $U:=\{U_{jn}\}$ with
$V_{jn}\in L^{\infty}(0,\ell_{jn})$ and coupling constants
$\alpha:=\{\alpha_j\in\R:\: j\in I\}$, we define the Schr\"odinger
operator $H_\alpha\equiv H_\alpha(\Gamma,U)$ by
   \begin{equation} \label{graph SO}
H_\alpha\{\psi_{jn}\} := \{\, -\psi''_{jn}+U_{jn}\psi_{jn}:\;
(j,n)\in I_\LL\,\}
   \end{equation}
on the domain consisting of all $\psi$ with $\psi_{jn}\in
W^{2,2}(0,\ell_{jn})$ satisfying suitable boundary conditions at
the vertices linking the boundary values
   \begin{equation} \label{boundary values}
\psi_{jn}(j):=\lim_{x\to 0+} \psi_{jn}(x)\,, \quad \psi'_{jn}(j):=
\lim_{x\to 0+} \psi'_{jn}(x)\,,
   \end{equation}
where the point $x=0$ is identified with $\XX_j$. Specifically, we
will work here with the so-called \emph{$\delta$ \em coupling:} at
any $\XX_j\in\VV$ we have $\,\psi_{jn}(j)=\psi_{jm}(j)=:\psi_j$
for all $n,m\in\nu(j)$, and
   \begin{equation} \label{delta}
\sum_{n\in\nu(j)} \psi'_{jn}(j)= \alpha_j\psi_j\,;
   \end{equation}
it is known that among all (non-trivial) boundary conditions which
make the operator (\ref{graph SO}) self-adjoint there are no other
with wavefunctions continuous at the vertices \cite{ES1}. The
particular case $\alpha=0$ represents the most simple boundary
conditions, called usually \emph{Kirchhoff} \cite{KS}, which we
will employ in the example of Sec.~\ref{Sinai}, however, for the
moment it is useful to consider the more general situation
(\ref{delta}). Furthermore, if the boundary $\BB\ne\emptyset$ we
assume Dirichlet boundary conditions there,
   \begin{equation} \label{free end bc}
\psi_j=0\,, \quad j\in I_\BB\,.
   \end{equation}
If $\Gamma$ is infinite one can look not only for bound states of
$H_\alpha$ but also for solutions of the equation
   \begin{equation} \label{local SO}
H_\alpha\psi= k^2\psi
   \end{equation}
referring to the continuous spectrum. To describe the generalized
eigenfunctions we consider in such a case the class $D_{\rm
loc}(H_\alpha)$ which is the subset in $\bigvee_{(j,n)\in I_\LL}
L^2(0,\ell_{jn})\,$ (the direct sum) consisting of the functions
which satisfy all the requirements imposed at $\psi\in
D(H_\alpha)$ except the global square integrability. The
conditions (\ref{delta}) define self-adjoint operators also if the
$\alpha_j$'s are formally put equal to infinity. We exclude this
possibility, which corresponds to \emph{Dirichlet decoupling} of
the operator at $\XX_j$ turning the vertex effectively into $N_j$
points of the boundary.

We need the decoupling, however, to state the result. Let
$H_\alpha^D$ be the operators obtained from $H_\alpha$ by changing
the conditions (\ref{delta}) at the points of $\II$ to Dirichlet
and denote $\KK:=\{ k:\, k^2\in \sigma(H_\alpha^D)\}$. In the
particular case when the particle is free at graph edges,
$U_{jn}=0$, this set is given explicitly as $\KK:=\{ \pi n
\ell_{jn}^{-1}: \, (j,n)\in\II_\LL,\, n\in\N_+\}$. We will adopt
several assumptions, namely
   \begin{description}
   \vspace{-.8ex}
\item{\em (i)$\:$} all the potentials of the family $\{U_{jn}\}$
are uniformly bounded for $(j,n)\in I_\LL\,$, \vspace{-.8ex}
\item{\em (ii)} $\ell_0:=\inf \{\,\ell_{jn}:\, (j,n)\in
I_\LL\}>0\,$, \vspace{-.8ex}
\item{\em (iii)} $L_0:=\sup \{\,\ell_{jn}:\, (j,n)\in
I_\LL\}<\infty\,$, \vspace{-.8ex}
\item{\em (iv)} $\;N_0:=\max \{\,\# \nu(j):\, j\in I\,\}
<\infty\,$.
   \end{description}

To formulate the result, we need a few more notions. On
$\LL_{nj}\equiv [0,\ell_{jn}]$, where the right endpoint
identified with the vertex $\XX_j$, we denote as $u_{jn}$ and
$v_{jn}$ the solution to $-f''+U_{jn}f=k^2f$ which satisfy the
normalized Dirichlet boundary conditions
$$
u_{jn}(\ell_{jn})= 1\!-\!(u_{jn})'(\ell_{jn})=0\,, \;\; v_{jn}(0)=
1\!-\!(v_{jn})'(0)=0\,;
$$
their Wronskian is naturally equal to $W_{jn}= -v_{jn}(\ell_{jn})
=u_{jn}(0)$. After this preliminary we can specify the result of
\cite{Ex1} to the present situation. \\ [.3em]
\emph{Theorem:} (a) Let assumptions (i)--(iv) be satisfied and
suppose that $\psi\in D_{loc}(H_\alpha)$ solves (\ref{local SO})
for some $k\not\in\KK_\alpha$ with $k^2\in\R$, $\im k\ge 0$. Then
the corresponding boundary values (\ref{boundary values}) satisfy
the equation
   \begin{equation} \label{discrete delta}
\sum_{n\in\nu(j)\cap I_\II} {\psi_n\over W_{jn}}\,-\, \left(\,
\sum_{n\in\nu(j)} {(v_{jn})'(\ell_{jn}) \over W_{jn}} -\alpha_j\,
\right)\psi_j\,=\, 0\,.
   \end{equation}
Conversely, any solution $\{\psi_j:\, j\in I_\II\}$ of the system
to (\ref{discrete delta}) determines a solution of (\ref{local
SO}) by
   \begin{eqnarray*}
\psi_{jn}(x)= {\psi_n\over W_{jn}}\,u_{jn}(x) -\,{\psi_j\over
W_{jn}}\,v_{jn}(x) \;\;
& {\rm if} & n\in \nu(j)\cap I_\II\,, \label{reconstruction i} \\
\psi_{jn}(x)= -\,{\psi_j\over W_{jn}}\,v_{jn}(x) \;\; & {\rm if} &
n\in \nu(j)\cap I_\BB\,. \label{reconstruction b}
   \end{eqnarray*}
(b) Under (i), (ii), $\psi\in L^2(\Gamma)$ implies that the
solution $\{\psi_j\}$ of the system (\ref{discrete delta}) belongs
to $\ell^2(I_\II)$.
\\ [.3em]
(c) The opposite implication is valid provided (iii), (iv) also
hold, and $k$ has a positive distance from from $\KK$.


\setcounter{equation}{0}
\section{Approximation by lattice graphs}

As the next step, let us inspect how the above duality looks under
simplifying assumption: we will suppose that (a) all the graph
edges have \emph{the same length} $\ell>0$ and (b) all the
potentials $U_{jn}$ vanish. Then the ``elementary'' solutions can
be made explicit,
 $$
 u_{jn}(x)= \frac{1}{k}\, \sin k(x-\ell)\,, \quad
 v_{jn}(x)= \frac{1}{k}\, \sin kx\,,
 $$
with the Wronskian $W_{jn}= -\frac{1}{k}\, \sin k\ell$, and the
dual system of equations (\ref{discrete delta}) becomes
   \begin{equation} \label{equilateral}
 -\sum_{n\in\nu(j)} \frac{\psi_n- \psi_j\cos k\ell}{ k^{-1}
 \sin k\ell}  +\alpha_j\psi_j= 0\,, \quad j\in I\,;
   \end{equation}
Notice that this is true even if some of the $\nu(j)$ correspond
to points of the boundary, because we assume Dirichlet condition
(\ref{free end bc}) there so the corresponding $\psi_n$'s are
zero.

So far we worked with the local metric on $\Gamma$, now we will
regard the graph as a subset in $\R^\nu$ and assume that the local
metric coincides with the global one obtained by this embedding.
We will not strive for a most general result and concentrate on an
important particular case of a \emph{cubic lattice graph}
$\CC^\nu\equiv \CC^\nu(\ell) \subset \R^\nu$ whose vertices are
points $\{x_j(\ell)=(j_1\ell,\dots,j_\nu\ell):\: j_i\in\Z\}$ while
the edges are segments connecting pairs of vertices in which
values of a single index $j_i$ differ by one.
\\ [.3em]
\emph{Theorem:} Let $V:\R^\nu\to\R$ be a smooth function with
$\nabla V$ bounded. Put $\alpha_j(\ell):= V(x_j)\ell$ and consider
the family of operators $H_\alpha(\CC^\nu(\ell),0)$ with $\ell>0$.
Suppose that for any fixed $\ell$ and  $k\in\R$, the family
$\{\psi_j^\ell\}$ solves the equation (\ref{equilateral}), and
define a step function $\psi_\ell:\,\R^\nu\to \C$ by
 $$
 \psi_\ell(x):= \psi_j^\ell \quad {\rm if} \;\; -\frac12\ell
 \le (x-x_j)_i < \frac12\ell\,.
 $$
Suppose that the family $\{\psi_\ell\}$ converges to a function
$\psi:\,\R^\nu\to \C$ as $\ell\to 0$ in the sense that the
quantities $\varepsilon_j(\ell):= \psi(x_j)-\psi_\ell(x_j)$ behave
as
   \begin{equation} \label{stepconv}
 \sum_{n\in\nu(j)} \left(\varepsilon_n(\ell) -\varepsilon_j(\ell)
 \right) =o(\ell^2)\,.
   \end{equation}
Then the limiting function solves the equation
   \begin{equation} \label{limeq}
 -\Delta\psi(x) + V(x)\psi(x) = \nu k^2\psi(x)\,.
   \end{equation}
\emph{Proof:} Let $f$ be a $C^2$-smooth function, using its Taylor
expansion to the second order we find
  \begin{eqnarray*}
 \lefteqn{\frac{f(x+\ell) -f(x-\ell) -2f(x)\cos k\ell}{\ell k^{-1} \sin
 k\ell}} \\ [.2em] && = \frac{2k}{\ell} f(x) \tan\frac{k\ell}{2} + f''(x)
 \frac{k\ell}{\sin k\ell} + o(\ell)\,,
  \end{eqnarray*}
so the right-hand side tends to $f''(x)+k^2f(x)$ as $\ell\to 0$;
in fact, the error is $\OO(\ell^2)$ provided $f\in C^3$. Applying
this result to the function $\psi$ with respect to each of the
$\nu$ variables and combining it with the fact the family
$\{\psi_\ell(x_j)\}$ solves the equation (\ref{equilateral}) we
find
 \begin{eqnarray*}
 \lefteqn{\Delta\psi(x_j) +\nu k^2\psi(x_j) - V(x_j)\psi(x_j)}
 \\ [.2em] && = \left(
 \frac{\ell}{k} \sin k\ell \right)^{-1} \sum_{n\in\nu(j)}
 \left(\varepsilon_n(\ell) -\varepsilon_j(\ell) \right)
 + o(\ell)
 \end{eqnarray*}
and the right-hand side tends to zero by assumption.

\vspace{.3em}

Let us add a few comments: \\ [.1em]
(a) The requirement $k\not\in\KK$ means no restriction here,
because for a fixed $k$ it is satisfies if $\ell$ is small enough. \\
[.1em]
(b) The rectangular lattice used to prove the theorem is not
substantial. The same argument can be used, e.g., to prove the
theorem for a graph resulting from tessellation of the plane by a
lattice of equilateral triangles. Recall also that for rectangular
lattices a similar result can be proven by a different method
using resolvent of the Hamiltonian - see \cite{melnikov}.
\\
[.1em]
(c) Notice that the limiting energy has to be rescaled to $\nu
k^2$, where $\nu$ is the dimension, roughly speaking because all
``local'' momentum components are equal. This claim remains valid
when the we replace a rectangular graph with a triangular one.
\\
[.1em]
(d) The fact that the motion on the graph is locally restricted to
particular directions only does not mean that on larger scales the
particle cannot move through such lattice in any possible angle in
a zig-zag way. To illustrate the last claim recall how Fermi
surface looks like on a 2D square lattice in the free case,
$\alpha_j=0$ for any $j\in I$. By \cite{EG} it is described by the
equation
 $$
 \cos\theta_1\ell + \cos\theta_2\ell =2\cos k\ell\,,
 $$
where $\theta_i$ are the quasimomentum components, thus for small
$\ell$ we have at the bottom of the spectrum
 $$
 2k^2 = \theta_1^2 +\theta_2^2 +\OO(\ell^2)\,,
 $$
which looks like the free ``continuous'' motion, apart of the
factor $2$ multiplying the energy.

A similar result can be derived if the lattice graphs do not cover
the whole $\R^\nu$. Consider an open set $\Omega\subset\R^\nu$ and
call $\CC^\nu_\Omega\equiv \CC^\nu_\Omega(\ell)$ the subgraph of
$\CC^\nu(\ell)$ whose vertices are all points $x_j$ contained in
$\Omega$. Let $\PP^\nu_\Omega(\ell)$ denote the union of all
closed hypercubes of $\CC^\nu_\Omega(\ell)$, i.e. the ``volume''
of such a lattice in $\R^\nu$. If an edge of
$\CC^\nu_\Omega(\ell)$ belongs to the boundary of
$\PP^\nu_\Omega(\ell)$ we delete it. It may also happen that
$\PP^\nu_\Omega$ is non-convex, i.e. there is an axis along which
a boundary point has neighbors in $\CC^\nu_\Omega(\ell)$ in both
directions, then we regard the corresponding vertex as a family of
disconnected vertices belonging to the boundary of
$\CC^\nu_\Omega(\ell)$; we call the lattice modified in this way
$\tilde\CC^\nu_\Omega(\ell)$. Mimicking the above argument, we
arrive at the following conclusion: \\ [.3em]
\emph{Theorem:} Suppose that the potential $V:\Omega\to\R$ is
smooth with $\nabla V$ bounded and set $\alpha_j(\ell):=
V(x_j)\ell$. Consider the dual system associated with the family
$\{H_\alpha(\tilde\CC^\nu_\Omega(\ell),0):\:\ell>0\}$ and its
solutions $\{\psi_j^\ell\}$. Under the same convergence assumption
as above, the limiting function $\psi$ solves the equation
   \begin{equation} \label{limeq}
 -\Delta\psi(x) + V(x)\psi(x) = \nu k^2\psi(x)
   \end{equation}
with Dirichlet condition, $\psi(x)=0$ for $x\in\partial\Omega$.

\vspace{.3em}

Let us stress an important feature, namely that the described
result has a \emph{local character}. It is especially important
from the viewpoint of the example discussed below, where we will
violate regularity of the solution at a fixed points by attaching
leads to $\Omega$. This means that the solution has a singularity
at such a junction, a logarithmic one for $\nu=2$, which enters
the coupling between the billiard and the lead. Outside the
connection points, however, the graph approximants do still
converge to solution of the appropriate Schr\"odinger equation.


\setcounter{equation}{0}
\section{Example: Sinai billiard graphs} \label{Sinai}

\begin{figure}[htb]
\begin{center}
\rotatebox{270}{\scalebox{1.6}{\includegraphics[height=5cm,
width=5cm]{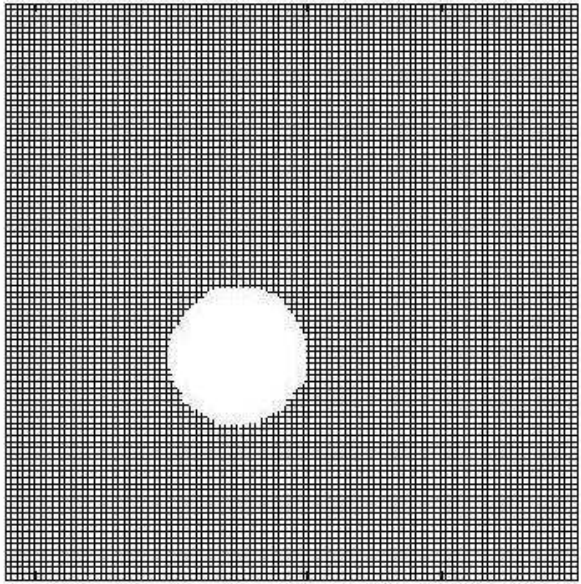}}}
\end{center}
\caption{Sinai billiard graph} \label{sinai_f}
\end{figure}

We will consider a rectangular $N\times N$ lattice graph with a
circular part removed reminiscent of a Sinai billiard which
according to the above result such a structure can approximate --
cf.~Fig.~\ref{sinai_f}. For practical calculations we choose
$N=97$ and $\alpha_j=0,\, U_{jn}=0$; at the graph boundary we
impose Dirichlet conditions. The lattice graph spacing is set to
be $\ell=0.15$.

\begin{figure}[htb]
\begin{center}
\rotatebox{0}{\scalebox{1.6}{\includegraphics[height=5.cm,
width=5cm]{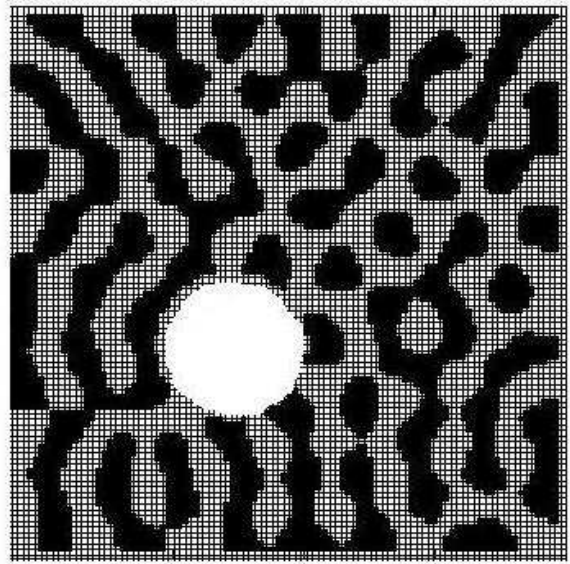}}}
    \end{center}
\caption{Nodal domains an eigenfunction} \label{nod_f}
\end{figure}

First we look at the nodal zone structure of one of the
eigenfunctions -- cf.~Fig.~\ref{nod_f}. The vertices of the graph
are marked as black when the value of the eigenfunction is
positive at the vertex or white when it is negative. What comes
out is a pattern similar to that of the hard-wall Schr\"odinger
problem on the corresponding Sinai billiard.

As we have indicated we want to compare transport properties of
such systems in the situation when an incoming and outgoing lead
is attached to the graph (at the points $(14,40)$ and $(59,80)$)
and to the billiard at the corresponding places. Adding a lead to
a graph, represents no problems: at the incoming / outgoing vertex
a semiinfinite is added and  the resulting five edges are again
coupled by Kirchhoff conditions, (\ref{delta}) with $\alpha_j=0$.
On the other hand, coupling a billiard to leads needs explanation.
Here we use a standard method and describe the attached leads
(attached antenna in the case of microwave billiards) by
Sommerfeld radiation boundary conditions. Using this approach the
attached lead is replaced by a small circle and the following
boundary conditions are imposed on its boundary:
 \begin{equation}
 {{\partial\psi}\over{\partial\overrightarrow{n}}}+ik\psi=2ik
 \label{radiat1}
 \end{equation}
 for the incoming lead and
\begin{equation}
 {{\partial\psi}\over{\partial\overrightarrow{n}}}+ik\psi=0
 \label{radiat2}
 \end{equation}
for the outgoing one. The radius of the circles is much smaller
then the length of graph bonds. We have used $r=0.01$ for the
numerical tests. Another possibility is to  relax the regularity
requirement to solution $\phi$ to a two-dimensional Helmholtz
equation at the junction points. This approach is formally
equivalent to the limit $r\to 0$ and is described in  \cite{ES2},
\cite{BG}, \cite{etv} and \cite{ES3}. However since we are
interested in global structures that extend over the whole graph
(billiard) the detail character of the connection is not
important.

\begin{figure}[htb]
\begin{center}
\rotatebox{-90}{\scalebox{1.2}{\includegraphics[height=7.5cm,
width=4.cm]{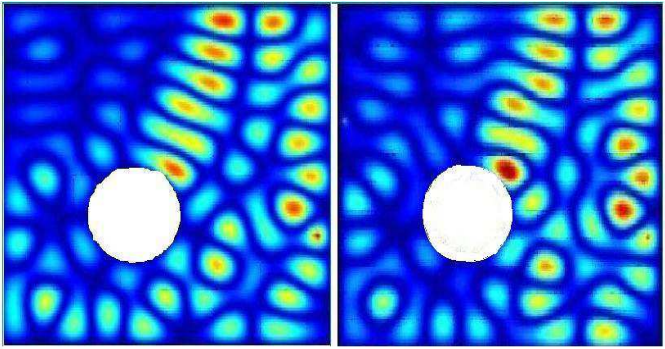}}}
\end{center}
\caption{Eigenfunction comparison in terms of probability density.
In the left picture squared modulus of the graph eigenfunction
corresponds to energy $E$ referring to the momentum $k=1.65$. The
right picture shows the same for billiard eigenfunction of energy
$2E$. The color scale ranges from zero (dark blue) to the maximum
value (dark red).} \label{compar_f}
\end{figure}

Let us start with comparing the wavefunctions on the graph with
those obtained for the corresponding two-dimensional billiard. A
typical result is displayed on the Figure \ref{compar_f} where we
have plotted the absolute value of the wavefunction. For the
graph, on the other hand, the values of the solution on the
vertices are shown.

Speaking of phase-related effects, a primary quantity of interest
is the \emph{probability current} which in (an open) billiard is
given conventionally by
 \begin{equation}
 \vec\jmath(\vec x) = \im \left( \bar\psi\nabla\psi \right)(\vec x)
\label{current}
\end{equation}

\begin{figure}[htb]
\begin{center}
\rotatebox{-90}{\scalebox{1.4} {\includegraphics[height=6.cm,
width=4.5cm]{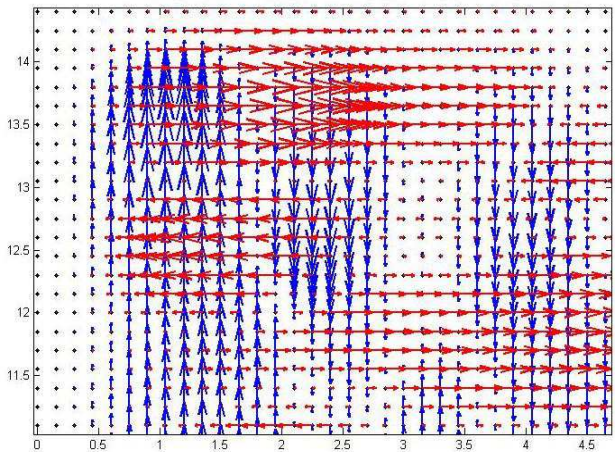}}}
\end{center}
\caption{Probability currents on $\Gamma$} \label{grafflow_f}
\end{figure}

\begin{figure}[htb]
\begin{center}
\rotatebox{-90}{\scalebox{1.6}{\includegraphics[height=5.5cm,
width=4.0cm]{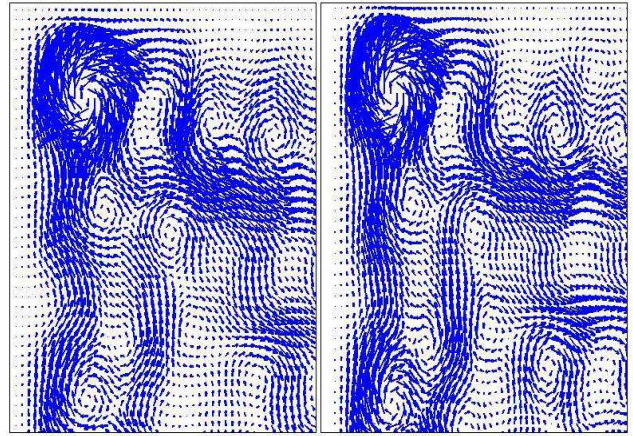}}}
\end{center}
\caption{Determining the vector field. The left current is
obtained by the vector summation while the current inside the
billiard is plotted on the right side} \label{flow}
\end{figure}

On the graph the current flows along the edges and has therefore a
prescribed direction -- cf.~Fig.~\ref{grafflow_f} -- so it is not
obvious what is the quantity to be compared to  the two
dimensional case. The natural possibility is to add the
``horizontal'' and ``vertical'' flows at each vertex  by a vector
summation and construct a in such a way a vector field. It turns
out that leads indeed to the correct probability current inside
the two dimensional billiard. We computed the currents using  the
above procedure are compared them with the current inside the two
dimensional billiard that was evaluated with the help of the
formula (\ref{current}). The result is plotted on the Figure
\ref{flow}.

In conclusion we have demonstrated the existence of global
structures on large graphs. The structures do not depend on the
local graph topology. We were able to prove that the large scale
structures are the same on a square graph, where each vertex
connects 4 bond, and on a graph consisting of equilateral
triangles when 6 bonds are connected at each vertex. Moreover
numerical results show that the structures do not change also for
other types of graphs (although a rigorous proof is missing).

The structures extend over hundreds of graph vertices. They make
up the manifestation of complex interference effects and are as
such difficult to understand. The way out is to employ embedding
of the graph into the appropriate ambient space and proving that
the graph wavefunction converges to a corresponding solution of
Schr\"odinger equation with the scaled energy.


\acknowledgements The research was supported in part by ASCR and
its Grant Agency within the projects IRP AV0Z10480505 and
A100480501.

\end{document}